\documentclass{elsart}

\usepackage{axodraw}
\usepackage{color}

\newcounter{bla}
\newenvironment{refnummer}{%
\list{[\arabic{bla}]}%
{\usecounter{bla}%
 \setlength{\itemindent}{0pt}%
 \setlength{\topsep}{0pt}%
 \setlength{\itemsep}{0pt}%
 \setlength{\labelsep}{2pt}%
 \setlength{\listparindent}{0pt}%
 \settowidth{\labelwidth}{[9]}%
 \setlength{\leftmargin}{\labelwidth}%
 \addtolength{\leftmargin}{\labelsep}%
 \setlength{\rightmargin}{0pt}}}
 {\endlist}

\begin{document}
\begin{frontmatter}
\begin{flushright}
SFB/CPP-08-53\\
TTP08-32
\end{flushright}

\title{Feynman Integral Evaluation by a Sector decomposiTion Approach (FIESTA)}

\author[a,b]{A.V.~Smirnov}, 
\author[b]{M.N.~Tentyukov}


\address[a]{
Scientific Research Computing Center - Moscow State University, 119991 Moscow, Russia}
\address[b]{
Institut f\"{u}r Theoretische Teilchenphysik - Universit\"{a}t
Karlsruhe, 76128 Karlsruhe, Germany}

\begin{abstract}

Up to the moment there are two
known algorithms of sector decomposition: an original private
algorithm of Binoth and Heinrich and an algorithm made public last
year by Bogner and Weinzierl.  We present a new program performing
the sector decomposition and integrating the expression afterwards.
The program takes a set of propagators and a set of indices as input
and returns the epsilon-expansion of the corresponding integral.

\begin{flushleft}

\end{flushleft}


\end{abstract}

\end{frontmatter}

{\bf PROGRAM SUMMARY}

\begin{small}
\noindent
{\em Manuscript Title:} Feynman integral evaluation by a sector decomposition approach\\
{\em Authors: }A.V.~Smirnov, M.N.~Tentyukov\\
{\em Program Title: } FIESTA \\
{\em Journal Reference:}\\
{\em Catalogue identifier:}\\
{\em Licensing provisions: } GPLv2\\
{\em Programming language: }{\tt Wolfram Mathematica} 6.0 and C\\
{\em Computer: } from a desktop PC to a supercomputer \\
{\em Operating system: }Unix, Windows\\
{\em RAM: }depends on the complexity of the problem  \\
{\em Number of processors used: } the program can successfully work with a single processor,
however it is ready to work in a parallel environment, and the use of multi-kernel
processor and multi-processor computers significantly speeds up the calculation 
{\em Classification:} 4.4 Feynman diagrams, 4.12  Other Numerical
Methods, 5 Computer Algebra, 6.5 Software including Parallel Algorithms\\
{\em External routines/libraries:} {\tt QLink} [1], {\tt Vegas} [2] \\
{\em Nature of problem:}
The sector decomposition approach to evaluating Feynman integrals
falls apart into the sector decomposition itself, where one
has to minimize the number of sectors; the pole resolution
and epsilon expansion; and the numerical integration of the resulting expression\\
{\em Solution method:}
The sector decomposition is based on a new strategy.
The sector decomposition, pole resolution and epsilon-expansion are
performed in {\tt Wolfram Mathematica} 6.0 [3].
The data is stored on hard disk via a special program, {\tt QLink} [1].
The expression for integration is passed to the C-part of the code,
that parses the string and performs the integration by the {\tt Vegas} algorithm [2].
This part of the evaluation is perfectly parallelized on multi-kernel
computers. \\
{\em Restrictions:} The complexity of the problem is mostly restricted
by CPU time required to perform the evaluation of the integral,
however there is currently a limit of maximum $11$
positive indices in the integral; this restriction
is to be removed in future versions of the code.\\
{\em Running time:} depends on the complexity of the problem\\
{\em References:}
\begin{refnummer}
\item http://qlink08.sourceforge.net, open source
\item G. P. Lepage, the Cornell preprint CLNS-80/447,1980.
\item http://www.wolfram.com/products/mathematica/index.html
\end{refnummer}
\end{small}

\newpage

\hspace{1pc}
{\bf LONG WRITE-UP}

\section{Introduction}

Sector decomposition in alpha (Feynman) parametric
representations of Feynman integrals originated as a tool for
analyzing the convergence
and proving theorems on renormalization and asymptotic expansions of Feynman integrals \cite{theory}.
The goal of this procedure is to
decompose the initial integration domain into appropriate subdomains (sectors)
and introduce, in each sector, new variables in such a way that the integrand
factorizes, i.e. becomes equal to a monomial in new variables
times a non-singular function. General mathematical results on Feynman
integrals exist up to now only in the case where all the external momenta are
Euclidean. However, in practice, one often deals with Feynman integrals on a
mass shell or at threshold.

Sector decomposition became a practical tool for evaluating Feynman
integrals numerically with the growth of computer power.  Practical
sector decomposition was introduced in \cite{BinothHeinric} and was
used to verify several analytical results for multiloop Feynman
integrals, including three-loop \cite{3box,3loop,4loop} results.  For
a good review on the sector decomposition we refer to
Ref.~\cite{Heinric}.

The first public algorithm has been published by Bogner and Weinzierl \cite{BognerWeinzierl}.
It proposes four strategies for the sector  decomposition; three of
those are guaranteed to terminate. Strategy A \cite{Zeillinger} is conceptually the simplest, however
it results in too many sectors; strategy B (Spivakovsky's strategy) has been described in \cite{Spivakovsky}, strategy C
has been created by Bogner and Weinzierl and is an improvement of strategy B.
Strategy X is likely to share the ideas of Binoth and Heinrich \cite{BinothHeinric}
It is not guaranteed to terminate but results in less sectors, than the other strategies.

This paper presents a new algorithm {\tt FIESTA} for evaluating Feynman integrals with the
use of sector decomposition.
We reproduced the strategies A, B and X in the code and presented a new strategy S,
based on original ideas. Our strategy is also guaranteed to terminate,
but results in more sectors than strategy X, however the difference is
between S and X is not so significant as with B and X.
We provide an example where the strategy X does not terminate.
For details and the table comparing the various strategies please refer to Appendix~\ref{A}.

Our algorithm can successfully work with a single processor,
however it is ready to work in a parallel environment, and the use of multi-kernel
processors and multi-processor computers significantly speeds up the calculation.

To benchmark the power of {\tt FIESTA}, we evaluated the triple box diagram
\cite{3box} (which starts from $\varepsilon^{-6}$ poles) up to the finite
part.  We used the strategy X and took the symmetries of the diagram
into account (see section \ref{syntax} for the syntax).  A Xeon double
processor computer was used (8 cores Intel(R) Xeon(R) CPU E5472,
3.00GHz, 1600FSB, 2x6MB L2 cache, 32 GB RAM, 4.6TB local hard drive).
The evaluation took less than $3$ days ($62.4$ hours); the job used
maximally 1GB of RAM and 40GB on hard drive. All the
answers are within the 1 percent error estimate from the existing
analytical result.

We tested {\tt FIESTA} only on Windows and Linux platforms but there are no
Linux-specifics; it should be possible to compile the sources on any Unix
platform. However, we provide the pre-compiled binary files only for
Windows on x86 and Linux on x86-64.

\section{Theoretical background}
\label{theory}

{\tt FIESTA} calculates Feynman integrals with the sector decomposition approach.
It is based on the $\alpha$-representation of Feynman integrals.
After performing Dirac and Lorentz algebra one is left with a scalar dimensionally regularized Feynman integral \cite{dimreg}
\begin{eqnarray}
  F(a_1,\ldots,a_n) &=&
  \int \cdots \int \frac{\mbox{d}^d k_1\ldots \mbox{d}^d k_l}
  {E_1^{a_1}\ldots E_n^{a_n}}\,,
  \label{FI}
\end{eqnarray}
where $d=4-2\varepsilon$ is the space-time dimension, $a_n$ are
indices, $l$ is the number of loops and $1/E_n$ are propagators. We
work in Minkowski space where
the standard propagators are the form $1/(-p^2+m^2-i0)$.
Other propagators are permitted, for example,
$1/(v\cdot k\pm i0)$ may appear in diagrams
contributing to static quark potentials or in HQET where $v$ is the quark velocity.
Substituting
\begin{eqnarray}
    \frac{1}{E_i^{a_i}}=\frac{e^{ai\pi/2}}{\Gamma(a)}\int_0^\infty \mbox{d}\alpha \alpha^{a_i-1} e^{-iE_i\alpha},
\label{AlphaSubstitution}
\end{eqnarray}
changing the integration order, performing the integration over loop momenta,
replacing $\alpha_i$ with $x_i \eta$ and integrating over $\eta$
one arrives at the following formula (see e.g. \cite{Smirnov}):
\begin{eqnarray}
\nonumber
    &&F(a_1,\ldots,a_n) =
   \\
    &&\frac{\Gamma(A-l d/2)}{\prod_{j=1}^n \Gamma(a_j)}
            \int_{x_j\geq 0} d x_i\ldots d x_{n} \delta\left(1-\sum_{i=1}^n x_i \right)
                \left(\prod_{j=1}^n x_j^{a_j-1}\right) \frac{U^{A-(l+1)d/2}}{F^{A-ld/2}},
\label{Alpha}
\end{eqnarray}
where $A=\sum_{i=1}^n a_n$ and
$U$ and $F$ are constructively defined polynomials of $x_i$.
The formula (\ref{Alpha}) has no sense if some of the indices are non-positive integers,
so in case of those the integration is performed according to the rule
\[
\int_{0}^{\infty}d x \frac{x^{(a-1)}}{\Gamma(a)} f(x) = f^{(n)} (0)
\]
where $a$ is a  non-positive integer.

After performing the decomposition of the integration region into the so-called
\textit{primary sectors} \cite{BinothHeinric} and making a variable replacement, one results
in a linear combination of integrals of the following form:
\begin{eqnarray}
 \int_{x_j=0}^1 d x_i\ldots d x_{n'}\left(\prod_{j=1}^{n'} x_j^{a_j-1}\right) \frac{U^{A-(l+1)d/2}}{F^{A-ld/2}}
\label{Cube}
\end{eqnarray}

If the functions $\frac{U^{A-(l+1)d/2}}{F^{A-ld/2}}$ had no singularities in $\varepsilon$,
one would be able to perform the expansion in $\varepsilon$ and perform the numerical integration
afterwards. However, in general one has to resolve
the singularities first, which  is not possible for general $U$ and
$F$. Thus, one
starts a process the sector decomposition aiming to end with a sum
of similar expressions, but with new functions $U$ and $F$ which have no singularities (all the singularities
are now due to the part $\prod_{j=1}^n {x'}_j^{a'_j-1}$). Obviously it is a good
idea to make the sector decomposition process constructive and to end
with a minimally possible number of sectors. The way sector decomposition is performed
is called a \textit{sector decomposition strategy} and is an essential
part of the algorithm.

After performing the sector decomposition one can resolve the singularities by
evaluating the first terms of the Taylor series:
in those terms one integration is taken analytically, and the remainder
has no singularities. Afterwards the $\varepsilon$-expansion can be performed
and finally one can do the numerical integration and return the result.

Please keep in mind
that this approach works only using numerical integration:
the values for all invariants should be substituted at the very early stage,
after generating the functions $U$ and $F$.

\section{Overview of the software structure}

{\tt FIESTA} is written in {\tt Wolfram Mathematica} 6.0 and C.
The user is not supposed to use the C part directly as it is
launched from {\tt Mathematica} via the Mathlink protocol.
The C part is performing the numerical integration
with the use of the {\tt Vegas} \cite{Vegas1,Vegas2} algorithm
implemented as a FORTRAN program.
The {\tt Mathematica} part can work independently,
however the C-integration is much more powerful than
the {\tt Mathematica} built-in one.
Hence to calculate complicated integrals one should use the C-integration.

\section{Description of the individual software components}

The {\tt Mathematica} part is a $.m$
file that is loaded into {\tt Mathematica} simply by {\tt <<FIESTA\_1.0.0.m}\footnote{
The paths to the {\tt CIntegrate} binary, to the {\tt QLink}\cite{QLink}
and the database should be specified beforehand; see section \ref{Inst} for details.}.

The {\tt Mathematica} part of the program takes as input the set of loop momenta,
the set of propagators,
the set of indices and the required order of $\varepsilon$-expansion to be evaluated.
It performs the differentiation (in case of negative indices),
the sector decomposition, the resolution of singularities, the $\varepsilon$-expansion
and prepares the expressions for the numerical integration.
The numerical integration can both be performed by {\tt Mathematica} or by
the C-part of {\tt FIESTA}. The {\tt Mathematica} integration has the advantage of
being able to work with complex numbers (in future version this feature
will also be added to the C-part), however, the C-integration is
much more powerful and should be used if one goes for complicated integrals.

The C-integration is called from {\tt Mathematica} via the Mathlink protocol.
This part of the algorithm can perfectly work in a parallel environment:
one simply has to specify the number of copies of {\tt CIntegrate} to be launched.
The {\tt Mathematica} part distributes the integration tasks between those
copies and collects the result, preparing the expression of the next
order at the same time.

The C-integration comes as a set of source files and should be
compiled first.
The binary files of
{\tt CIntegrate} for Linux and Windows can be downloaded from the web-page of the project
\\
http://www-ttp.particle.uni-karlsruhe.de/$\sim$asmirnov/FIESTA!.htm

The C-integration is using {\tt Vegas} \cite{Vegas1,Vegas2} to
perform the integration.  It takes a string with the expression from
{\tt Mathematica} as input, and performs an internal compilation of
this string to be able to evaluate the expression fast afterwards. No
external compiler is required and no files are used for data exchange.

In complicated cases one should use the {\tt QLink} program
\cite{QLink} to store intermediatedata from {\tt Mathematica} on disk instead of
RAM.  {\tt QLink} can be downloaded as a binary or as a source (it
comes under the terms of GNU GPLv2).

\section{Installation instructions}
\label{Inst}
In order to install {\tt FIESTA}, the user has to download the installation
package from the following URL:
\\
http://www-ttp.particle.uni-karlsruhe.de/$\sim$asmirnov/data/FIESTA.tar.gz,
\\
unpack it and follow the instructions in the file INSTALL.

The {\tt Mathematica} part of {\tt FIESTA} requires almost no installation,
one only needs to copy the {\tt FIESTA\_1.0.0.m} file and edit
the default paths QLinkPath, CIntegratePath and DataPath
in this file, e.g.:
\begin{itemize}
\item
{\tt QLinkPath="/home/user/QLink/QLink"};
\item
{\tt CIntegratePath="/home/user/FIESTA/CIntegrate"};
\item
{\tt DataPath="/tmp/user/temp"};
\end{itemize}
Here \verb|QLinkPath| is a  path to the executable {\tt QLink} file,
\verb|CIntegratePath| is a  path to the executable {\tt CIntegrate}
file, and \verb|DataPath| is a path to the database directory.
For the Windows system, these paths should look like
\begin{itemize}
\item
{\tt QLinkPath="C:/programs/QLink/QLink.exe"}\footnote{{\tt Mathematica} uses normal slashes for paths both in Unix and Windows.};
\item
{\tt CIntegratePath="C:/programs/FIESTA/CIntegrate.exe"};
\item
{\tt DataPath="D:/temp"};
\end{itemize}
Note, the program will create a big IO traffic to the directory
\verb|DataPath|, thus it is better to put this directory on a fast
local disk.

Alternatively, one can specify all these paths manually after loading
the file {\tt FIESTA\_1.0.0.m} into {\tt Mathematica}.

Please note that the code requires {\tt Wolfram Mathematica} 6.0 to be installed
and will not work correctly under lower versions of {\tt Mathematica}.

In order to work with nontrivial integrals, the user must install
{\tt QLink} and the C-part of {\tt FIESTA}, the {\tt CIntegrate} program. The {\tt
  QLink}
\cite{QLink} can be downloaded as a binary file or compiled from the
sources. If the user decides to use pre-compiled {\tt CIntegrate} executable
file, he has to place the file to some location and edit the paths
in the file {\tt FIESTA\_1.0.0.m} as it is described above. If the
user wants to compile the executable file himself he must have the
Mathematica Developer Kit installed on his computer.

Under Unix, the user must edit the file {\tt comdef.h} to be sure that the
macro WIN is not defined, i.e., comment the
macrodefinition \verb|#define WIN 1|. Then the self-explanatory file
{\tt Makefile} can be used by means of running the ``make'' command
(the path to the Mathematica Developer Kit libraries should be provided). As a
result, the binary file {\tt CIntegrate} appears.

For the Windows installation, the user must edit the file {\tt comdef.h} to
ensure that the macro WIN is defined, \verb|#define WIN 1|.
The Windows makefile is provided in the package;
it is named ``{\tt CIntegrate.mak}'' and is used with the ``nmake /f CIntegrate.mak'';
To run this command one has to have the Microsoft Visual C++ installed
(an express edition of Microsoft Visual C++ can be downloaded for free from the Microsoft site).
As an addition to the programs and packages mentioned above,
the user should install the Microsoft Platform Software Developer Kit
(It can also  be downloaded freely after validating a genuine copy of Microsoft Windows)
and the Intel Fortran compiler (an evaluation copy is also available
form the Intel site).

\section{Test run description}
\label{syntax}

To run {\tt FIESTA} load the {\tt FIESTA\_1.0.0.m} into {\tt Wolfram Mathematica} 6.0.
To evaluate a Feynman integral one has to use the command

{\tt SDEvaluate[\{U,F,l\},indices,order]},

where {\tt U} and {\tt F} are the functions from formula
(\ref{Alpha}), {\tt l} is the number of loops,
{\tt indices} is the set of indices and {\tt order} is the required
order of the $\varepsilon$-expansion.

To avoid manual construction of $U$ and $F$ one can use a build-in function {\tt UF}
and launch the evaluation as

{\tt SDEvaluate[UF[loop\_momenta,propagators,subst],indices,order]},

where {\tt subst} is a set of substitutions for external momenta, masses and
other values (note, the code performs numerical integrations. Thus the
functions {\tt U} and {\tt F}
should not depend on any external values).

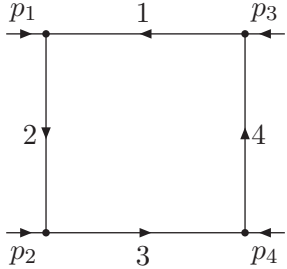
\begin{figure}[ht]
\begin{center}
\fcolorbox{white}{white}{
  \begin{picture}(135,125) (45,-5)
    \SetWidth{0.5}
    \SetColor{Black}
    \ArrowLine(75,95)(75,20)
    \ArrowLine(75,20)(150,20)
    \ArrowLine(150,20)(150,95)
    \ArrowLine(150,95)(75,95)
    \Vertex(75,20){1.41}
    \Vertex(75,95){1.41}
    \Vertex(150,95){1.41}
    \Vertex(150,20){1.41}
    \ArrowLine(165,95)(150,95)
    \ArrowLine(165,20)(150,20)
    \ArrowLine(60,95)(75,95)
    \ArrowLine(60,20)(75,20)
    \Text(150,15)[lt]{\small{\Black{$\;p_4$}}}
    \Text(75,15)[rt]{\small{\Black{$p_2\;$}}}
    \Text(112,15)[ct]{\small{\Black{$3$}}}
    \Text(75,57)[rc]{\small{\Black{$2\;$}}}
    \Text(150,57)[lc]{\small{\Black{$\;4$}}}
    \Text(150,100)[lb]{\small{\Black{$\;p_3$}}}
    \Text(75,100)[rb]{\small{\Black{$p_1\;$}}}
    \Text(112,100)[cb]{\small{\Black{$1$}}}
  \end{picture}
}
\end{center}
\caption{
\label{box}
The massless on-shell box diagram.
}
\end{figure}

Example:

{\tt SDEvaluate[UF[\{k\},\{-k$^2$,-(k+p$_1$)$^2$,-(k+p$_1$+p$_2$)$^2$,-(k+p$_1$+p$_2$+p$_4$)$^2$\},
\\
\{p$_1^2\rightarrow$0,p$_2^2\rightarrow$0,p$_4^2\rightarrow$0,
p$_1$ p$_2\rightarrow$-S/2,p$_2$ p$_4\rightarrow$-T/2,p$_1$ p$_4\rightarrow$(S+T)/2,
\\
S$\rightarrow$3,T$\rightarrow1$\}],
\{1,1,1,1\},0]
}

performs the evaluation of the massless on-shell box diagram
(Fig.~\ref{box}) where the Maldestam variables are equal to  $S=3$ and $T=1$.
A complete log of this example follows in Appendix~\ref{BB}.

To clear the results from memory use the {\tt ClearResults[]} command.

The code has the following options:

\begin{itemize}
\item {\tt UsingC}: specifies whether the C-integration should be used; default value: {\tt True};
\item {\tt CIntegratePath}: path to the {\tt CIntegrate} binary;
\item {\tt NumberOfLinks}: number of the {\tt CIntegrate} programs to be launched; default value: $1$;
\item {\tt UsingQLink}: specifies whether {\tt QLink} should be used to store data on disk;
works only with {\tt UsingC=True}; default value: {\tt True}; please
refer to Appendix~\ref{C} on the use of {\tt QLink};
\item {\tt QLinkPath}: path to the {\tt QLink} binary;
\item {\tt DataPath}: path to the place where {\tt QLink} stores the data;
for example, if {\tt DataPath=/temp/temp}, then the
code three directories: {\tt /temp/temp1},
{\tt /temp/temp2} and {\tt /temp/temp3}; those directories will be erased if existent;
the directory {\tt /temp} should exist;
\item {\tt PrepareStringsWhileIntegrating}: specifies whether the code should prepare expressions
of next order in epsilon at the same time with integrating; works only with {\tt UsingC=True} and {\tt UsingQLink=True};
should be set to {\tt False} on slow machines; default value: {\tt True};
\item {\tt IntegrationCut}: the actual low boundary of the integration domain (instead of zero); default value: $0$;
\item {\tt IfCut} and {\tt VarExpansionDegree}: if {\tt IfCut} is nonzero then the expression
is expanded up to order {\tt VarExpansionDegree} over some of the integration variables; the integration
function is evaluated exactly if the integration variable is greater that {\tt IfCut},
otherwise the expansion is taken instead; the default value of {\tt IfCut} is zero,
the default value of {\tt VarExpansionDegree} is $1$;
\item {\tt ForceMixingOfSectors}: if set to {\tt True}, forces the code to perform one integration
only for a given order of $\varepsilon$; default value is {\tt False};
please refer to
Appendix~\ref{F};

\item {\tt PrimarySectorCoefficients}: The usage of this option allows
  to take the symmetries of the diagram into account. If the diagram has symmetries,
then the primary sectors corresponding to symmetrical lines result in equal
contributions to the integration result.
Hence it makes sense to speed up the calculation by specifying the
coefficients before the primary sector integrands. For example, if two lines in the
diagram are symmetrical, one can have a zero coefficient before one of those
and $2$ before the second. {\tt PrimarySectorCoefficients} defines those
coefficients if set; the size of this list should be equal to the number of primary sectors;
\item {\tt STRATEGY}: defines which sector decomposition strategy is used;
{\tt STRATEGY\_0} is not exactly a strategy, but an instruction not to perform the sector decomposition;
{\tt STRATEGY\_A} and {\tt STRATEGY\_B} are the two strategies from Ref.~\cite{BognerWeinzierl} guaranteed to terminate;
{\tt STRATEGY\_S} (default value) is our strategy, producing better results than the preceding ones;
{\tt STRATEGY\_X} is an heuristic strategy from \cite{BognerWeinzierl};
likely to share the ideas of Binoth and Heinrich \cite{BinothHeinric}: powerful but not guaranteed to terminate;
\item {\tt ResolveNegativeTerms}: defines whether an attempt to get rid of negative terms in F should be performed;
default value: {\tt True}; see Appendix~\ref{B};
\item {\tt VegasSettings}: a list of $4$ numbers $\{p_1,r_1,p_2,r_2\}$, instructs the code to
evaluate the integral $r_1$ times with $p_1$ sampling points for warm-up and
 $r_2$ times with $p_2$ sampling points for the final integration;
 default value: $\{10000,5,100000,15\}$.
\end{itemize}

The code sometimes returns {\tt Indeterminate} as a result.
There are the following reasons possible:
\begin{itemize}
\item \textit{Negative function $F$ provided};

This might happen if the propagators in the input for {\tt UF} are of the form $-1/(-p^2+m^2-i0)$
instead of $1/(-p^2+m^2-i0)$;

\textit{Solution:} change the sign of the propagators;

\item \textit{Complex number as an answer};

If the integrand has complex values, then the C integration
cannot be used with the current version of {\tt FIESTA}.
If complex numbers are explicitly encountered in the integrand,
the integration will produce a syntax error.
Otherwise, if the expressions like $\log(-1)$ are encountered in the integrand,
it the integration program returns {\tt Indeterminate};

\textit{Solution:} One can try to use the {\tt UsingC=False} option,
however the built-in {\tt Mathematica} integration cannot work
in complicated examples and we don't guarantee correct results
if complex numbers are used. The support for complex numbers
will be provided in future versions of {\tt FIESTA};

\item \textit{Special singularities};

The standard sector decomposition approach works only for singularities for small values of variables.
Our code can resolve some extra singularities (see Appendix~\ref{B}),
but surely all types of singularities could not be be covered.

\textit{Solution:} One can try different values of the {\tt ResolveNegativeTerms} option.
If the singularity is of a special type, one can try contact the
authors in order to try to make the resolution of singularities of this type automatic.

\item \textit{Numerical instability};

If all the singularities are for small values of integration variables, they are resolved in the code.
However, the numerical integration may experience problems with small values of variables (see Appendix~\ref{D}).

\textit{Solution:} One should use the {\tt IntegrationCut} or {\tt IfCut} together with {\tt VarExpansionDegree} options.
See Appendix~\ref{D} for details.

\end{itemize}

\section{Acknowledgements}

The authors would like to thank V.A.~Smirnov and M.~Steinhauser for
posing the problem of sector decomposition and for testing in on
various examples, to P.~Uwer for the consultation on {\tt Vegas}
and numerical integration and to G.~Heinrich for
useful discussions.

This work was supported in part by DFG through SBF/TR 9 and RFBR 08-02-01451-a.

\appendix
\section*{Appendix}

\section{Sector decomposition strategies}
\label{A}
The sector decomposition goal may be reformulated the following way:
suppose we have a polynomial $P$ of $n$ variables and a unitary hypercube $\mathcal U$ as integration area.
The task is to separate the integration region into a number of sectors so that
each of those sectors after a variable replacement turns into a unitary cube and
$P$ in those variables turns into a product of a monomial $M$ and a polynomial of
\textit{a proper form}, i.e. having $1$ as one of the summands.
This requirement is often sufficient for the polynomial $P$ to have no
zeros apart from the ones due to $M$
(this is valid for a wide class of integrals including those
with the same sign for all kinematic invariants).

The sector decomposition is usually performed recursively; for a given polynomial
$P$ one has an algorithm to define how to perform a single sector decomposition step.
Such an algorithm is also called \textit{a sector decomposition strategy}.

All known strategies separate $\mathcal U$ into parts the following way:
a subset $I=\{i_1,\ldots i_m\}$ of $\{1,\ldots,n\}$ is chosen and $m$ sectors are defined by
\[S_l=\{(x_1,\ldots,x_n)\in{\mathcal U} | x_{i_l}\geq x_{i_k} \forall
i_k\in I\},\]
($l=1\ldots m$).
The variable replacement in $S_l$ is defined by
\[ x_i = x'_i \;\forall i\not\in I\]
\[ x_{i_l} = x'_{i_l}\]
\[ x_{i_k} = x'_{i_l} x'_{i_k} \;\forall i\in I,k\neq l\]
It is easy to verify that the integration region in the new variables $x'$
is again a unitary cube.

We are extending the set of possible regions and variable replacements the following way:
for any vector $(v_1,\ldots, v_n)$ in the positive quadrant with at least two
non-zero coordinates we consider the set $I=\{i|v_i\neq 0\}$ and separate $\mathcal U$
into $m$ parts by
\[S_l=\{(x_1,\ldots,x_n)\in{\mathcal U} | x^{a_{i_l}}_{i_l}\geq x^{a_{i_k}}_{i_k} \forall i_k\in I\},\]
where $\{i_1,\ldots i_m\}=I$ and the exponents $a_i$ are defined by
\begin{equation}
\left(
\begin{array}{c}
a_{i_1}\\a_{i_2}\\a_{i_3}\\\vdots\\a_{i_m}
\end{array}
\right)=
\left(
\begin{array}{ccccc}
  0 & 1 & 1 & \ldots & 1 \\
  1 & 0 & 1 & \ldots & 1 \\
  1 & 1 & 0 & \ldots & 1 \\
  \vdots & \vdots & \vdots & & \vdots \\
  1 & 1 & 1 & \ldots & 0
\end{array}
\right)
\begin{array}{c}
-1\\ \\ \\ \\ \\ \\
\end{array}
\;
\left(
\begin{array}{c}
v_{i_1}\\v_{i_2}\\v_{i_3}\\\vdots\\v_{i_m}
\end{array}
\right)
\end{equation}
The variable replacement in $S_l$ is defined by
\[ x_i = x'_i \forall i\not\in I\]
\[ x_{i_l} = x'^{v_{i_l}}_{i_l}\]
\[ x_{i_k} = x'^{v_{i_k}}_{i_l} x'_{i_k} \;\forall i\in I,k\neq l.\]
It is only a little bit more complicated to verify that the
integration region in the variables $x'$ is still the unite hypercube.

Now let us describe how we choose the vector $v$.
We consider the set of weights $W$ of the polynomial $P$ defined as the set
of all possible $(a_1,\ldots,a_n)$ where $c x_1^{a_1}\ldots x_n^{a_n}$ is
one of the monomials of $P$. We will say that a weight is higher than another one
if their difference is a set of non-negative numbers.
If $P$ had a unique lowest weight, a monomial could be factorized out and
we would represent $P$ in the required form.
Hence it becomes reasonable to try to minimize the number of lowest weights
of $P$. We consider the convex hull of $W$ and choose one of its facets $F$
visible from the origin. Now $v$ is chosen to be the normal vector to $F$.

The reason of choosing such a vector is rather simple:
the vectors formed by the weights of the vectors $x'_{i_k}$ for $k\neq l$ are
orthogonal to $v$, therefore are lying in the facet $F$. Hence there is a good chance
that after a single sector decomposition step only one of the vertices of $F$
is left to be a lowest weight.

Such a sector decomposition step is guaranteed not to increase the norm
defined in the strategy A of \cite{BognerWeinzierl} (the Zeillinger strategy \cite{Zeillinger}).
If we can't find a facet such that the corresponding step decreases
the norm, we fall back to the strategy A and perform one step.
Hence our strategy is also guaranteed to terminate.
Practice shows that the strategy A steps are applied in no more than in five
percent cases.

The following table shows the number of sectors produced by different strategies on various examples:

\begin{tabular}{|c|c|c|c|c|c|}
  \hline
  Diagram & A &  B  & C & S & X \\
  \hline
  Box & 12 & 12 & 12 & 12 & 12 \\
  Double box & 755 & 586 & 586 & 362 & 293 \\
  Triple box & M & 114256 & 114256 & 22657 & 10155 \\
      D420   & 8898  &   564     &    564    &   180    &   F \\
  \hline
\end{tabular}

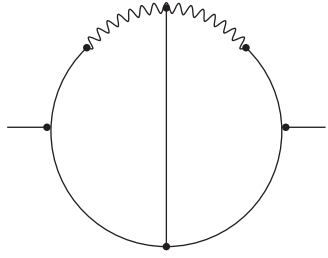
\begin{figure}[ht]
\begin{center}
\fcolorbox{white}{white}{
  \begin{picture}(120,93) (45,-42)
    \SetWidth{0.5}
    \SetColor{Black}
    \CArc(105,1.5)(43.5,133.6,406.4)
    \PhotonArc(105,10.5)(37.5,36.87,143.13){-2}{15.5}
    \Vertex(60,3){1.41}
    \Vertex(75,33){1.41}
    \Vertex(135,33){1.41}
    \Vertex(150,3){1.41}
    \Vertex(105,-42){1.41}
    \Line(105,-42)(105,48)
    \Line(165,3)(150,3)
    \Line(60,3)(45,3)
    \Vertex(105,48){1.41}
  \end{picture}
}
\end{center}
\caption{
\label{QPot}
 A diagram contributing to the 2-loop static quark potential
}
\end{figure}

D420 is a diagram contributing to the 2-loop static quark potential on Fig.~\ref{QPot} with the following
set of propagators: $\{-k^2,-(k-q)^2,-l^2,-(l-q)^2,-(k-l)^2,-v k,-v l\}$,
where $k$ and $l$ are loop momenta, $q^2=-1$, $qv=0$ and $v^2=1$.

``F'' means that the sector decomposition fails (we suppose an infinite loop).
``M'' means that the memory overflow happened during the sector decomposition on a 8Gb machine.
We suspect D420 to be a counterexample to the {\tt STRATEGY\_X}

Hence we recommend leaving the default option {\tt STRATEGY=STRATEGY\_S} (our strategy)
or experimenting with {\tt STRATEGY=STRATEGY\_X}
(based on the ideas of Binoth and Heinrich \cite{BinothHeinric},
not guaranteed to terminate).

\section{An example}
\label{BB}
Here we provide the listing of the {\tt Mathematica} session for
evaluation of the massless on-shell box diagram with $S=3$ and $T=1$ (we split some long lines).
{\small
\begin{verbatim}
In[1]:=  << FIESTA_1.0.0.m
FIESTA, version 1.0.0

In[2]:= UsingQLink=False;

In[3]:= SDEvaluate[UF[{k},{-k^2,-(k+p1)^2,-(k+p1+p2)^2,
        -(k+p1+p2+p4)^2},{p1^2->0,p2^2->0,p4^2->0,
        p1 p2->-S/2,p2 p4->-T/2,p1 p4->(S+T)/2,S->3,
        T->1}],{1,1,1,1},0]
External integration ready! Use CIntegrate to perform calls
FIESTA, version 1.0.0
UsingC: True
NumberOfLinks:1
UsingQLink: False
IntegrationCut: 0
IfCut: 0.
Strategy: STRATEGY_S
Integration has to be performed up to order 0
Sector decomposition..........0.0138 seconds; 12 sectors.
Variable substitution..........0.0055 seconds.
Decomposing ep-independent term..........0.0025 seconds
Pole resolution..........0.0081 seconds; 40 terms.
Expression construction..........0.0021 seconds.
Replacing variables..........0.0063 seconds.
Epsilon expansion..........0.0123 seconds.
Expanding..........0.0002 seconds.
Counting variables: 0.0002 seconds.
Preparing integration string..........0.0004 seconds.
Terms of order -2: 8 (1-fold integrals).
Numerical integration: 8 parts; 1 links;
Integrating..........0.106322 seconds; returned answer:
                                                1.333333
Integration of order -2: 1.333333
(1.333333)/ep^2
Expanding..........0.0005 seconds.
Counting variables: 0.0005 seconds.
Preparing integration string..........0.001 seconds.
Terms of order -1: 28 (2-fold integrals).
Numerical integration: 12 parts; 1 links;
Integrating..........0.409080 seconds; returned answer:
                                     -2.065743 +- 5.*^-6
Integration of order -1: -2.065743 +- 5.*^-6
(1.333333)/ep^2 + (-0.73241 +- 5.*^-6)/ep
Expanding..........0.0008 seconds.
Counting variables: 0.0009 seconds.
Preparing integration string..........0.0022 seconds.
Terms of order 0: 40 (2-fold integrals).
Numerical integration: 12 parts; 1 links;
Integrating..........0.862786 seconds; returned answer:
                                   -3.417375 +- 0.000012
Integration of order 0: -3.417375 +- 0.000012
(1.333333)/ep^2 + (-0.73241 +- 5.*^-6)/ep +
                                   (-4.386496 +- 0.000013)
Total time used: 1.52991 seconds.
                                            -6
                  1.33333   -0.73241 + 5. 10   pm46
Out[3]= -4.3865 + ------- + -----------------------
                      2               ep
                    ep

        + 0.000013 pm47
In[4]:= Quit
\end{verbatim}
}

The analytical answer for this integral is
\[
\frac{4}{3 \varepsilon^2}-\frac{2 \log(3)}{3 \varepsilon}-\frac{4 \pi ^2}{9}
\]
which is equal to
\[
1.3333333\frac{1}{\varepsilon^2} - 0.7324081\frac{1}{\varepsilon} -4.3864908.
\]

\section{Dealing with negative terms in $F$}

\label{B}
{\tt FIESTA} has been used to verify master integrals for the three-loop static
quark potential \cite{SSS}.
A typical phenomenon when evaluating those integrals
is that the function $F$ can involve terms like $(x_i-x_j)^2= x_i^2 - 2 x_i x_j + x_j^2 $.
As a result, not all monomials of $F$ have positive signs,
resulting in extra singularities for the integration.

To overcome those problems we added an additional feature to {\tt FIESTA}:
an attempt to get rid of
negative summands in $F$ before performing the sector decomposition
(those summands may lead to additional singularities that are not
treated properly by the basic approach).
The code tries to find two variables, $x_i$ and $x_j$, split the unitary cube into
two regions with $x_i\leq x_j$ and $x_i\geq x_j$ and do a variable replacement
(e.g. for $x_i\leq x_j$ we make the replacement $x'_i=x_j-x_i$).
In case of success, we result in a new function $F$ without monomials
with negative sign.

The ideas of this type are not original; it is a common situation when
certain classes of integrals require a special treatment,
in the papers of Binoth and Heinrich \cite{BinothHeinric} one
can find various examples where special operations
preceding the sector decomposition are performed. Still we consider
our automatic approach to the problem to be original: one does not require
to specify the singularities of this type for the code to resolve them.

\section{Avoiding memory limits in {\tt Mathematica} }
\label{C}
{\tt Wolfram Mathematica}, being a powerful CAS\footnote{Computer Algebra
  System} is widely used among researchers nowadays.
However, a significant drawback of the {\tt Mathematica} usage is the lack of RAM that is often encountered.
{\tt FIESTA} also had memory problems at the $\varepsilon$-expansion stage: an expression for a single
sector can become huge after the expansion, and in complicated examples one
can have many thousands of sectors. To avoid memory problems we used
  the {\tt QLink} program
to store data on hard disk.

After performing the sector decomposition we store all
the expressions in a database via {\tt QLink}. Afterwards the code never loads the whole set of expressions
into RAM. On the contrary, the code works with sectors one by one:
it loads an expression from the database, performs an operation with it and writes it back to the database.
The operation might be one of the following:
singularity resolution, $\varepsilon$-expansion, applying the {\tt Expand} function or
applying ``if conditions'' (see appendix \ref{D}).

It is worth noting that the $\varepsilon$-expansion does not use the {\tt Mathematica} {\tt Series} function
but explicit differentiation instead. This  has two reasons: first of all,
differentiation with subsequent expansion  works faster in {\tt Mathematica} 6.0
for the expressions we consider (remember that {\tt FIESTA}  uses {\tt
  Mathematica} 6.0
because of the powerful build-in integration
methods that were absent in 5.2). The second reason is to perform differentiation
at the very beginning but to wait with applying the {\tt Expand} function
(which takes more time than the differentiation!).
The latter function can be applied simultaneously with the integration
at a later stage: we prepare expressions of order $o+1$ while integrating expressions of order $o$.

Such an approach allows to keep the memory usage by {\tt Mathematica} minimal and to leave most
of the memory to the {\tt CIntegrate} copies.

\section{Resolution of singularities and its influence on the convergence of numerical integration}
\label{D}
As it has been noted in section \ref{theory}, the integrals
\begin{eqnarray}
 \int_{x_j=0}^1 d x_i\ldots d x_{n}(\prod_{j=1}^n x_j^{a_j-1}) Z,
\label{NoSing}
\end{eqnarray}
where $Z$ has no singularities might have singularities because of non-positive $a_j$.
Let us assume that $a_i\leq0$ and treat the integrand as a function
$x^{a_i-1} Y(a_i)$
with coefficients being polynomials of other variables. For readability we will
omit the index $i$. We replace $x^{a-1} Y(a)$ by
\begin{eqnarray}
Y(0)x^{a-1}+Y'(0)x^{a}+\ldots+\frac{1}{a!}Y^{(-a)}(0)x^{-1}+
\nonumber\\\nonumber
x^{a-1} (Y(a)-Y(0)-Y'(0)x-\ldots-\frac{1}{a!}Y^{(-a)}(0))
\end{eqnarray}
The items in the first line of the expression can be analytically integrated
over $x$ leaving us with one integration less. As for the expression
$R(a)=x^{a-1} (Y(a)-Y(0)-Y'(0)x-\ldots-\frac{1}{a!}Y^{(-a)}(0))$, it is known to have no
singularities at $x=0$, however it might result in
numerical instabilities at a later stage.

After performing the resolution of singularities and the $\varepsilon$-expansion,
one obtains a set of integrands depending only on integration variables over the
unit hypercube. The integrands arise from the expressions like $R$ (or even more
complicated ones in case of several
integration variables).

Those integrands have no singularities, so one could expect that the integration is performed
with ease. However, one can see that for small $x$ the $x^{a-1}$ is huge and the remainder
is small. Hence the remainder can't be evaluated exactly because it is a difference
of not-so-small numbers stored with a finite number of digits.

The first step to the solution is to expand the expressions before evaluating.
It leaves similar problems (a difference of huge numbers resulting in a normal number),
but with a smaller numerical instability. Still, in highly degenerate examples
the problem persists and prevents the code from evaluating the function
when the variables turn small.

{\tt FIESTA} has two methods to deal with the numerical instability.
The first method is the usage of the {\tt IntegrationCut} option.
This option sets the low integration limit to {\tt IntegrationCut} (instead of zero).
This approach may lead to high error estimate, so there is a more advanced method
(slower but providing better results).

The idea of the second method is to expand the integration
function over the integration variables around zero up to a certain order
(not all integration variables require the expansion,
but only the ones with a negative exponent in the preceding monomial).
Then one can integrate the original function for values greater than
some fixed number and the expanded one for smaller values.
The code is instructed to work this way by
the {\tt IfCut} and {\tt VarExpansionDegree} options.
Please note that even for {\tt IfCut=$10^{-2}$} and {\tt VarExpansionDegree=1}
one results in an $10^{-6}$ relative error estimate for the remainder that
is smaller than the integration error estimate for complicated examples.

There is one more technical problem in the second approach: the usage of the {\tt Mathematica} {\tt Series}
command can be quite slow. Hence to expand the function we need to substitute $x=0$
into the function and its derivatives. The direct substitution is impossible
for the same numerical reasons. The solutions comes with finding the minimal possible $s$
such that $(x^s f) (0)$ can be evaluated numerically. Now we can define a new function
$g = \frac{(x^s f)^{(s)}}{s!}$ and use the fact that $f^{(l)}(0)=C^{l+s}_s g^{(l)}(0)$
to expand the function.

Please note that the error estimate in the output is the one that
comes from the integration; if the cuts are used, one should keep in
mind that the real error might be greater for the reason that a
function different from the original one is integrated.

\section{External integration}
\label{E}

\subsection{Overview}
For the integration we use the well-known {\tt Vegas} algorithm
\cite{Vegas1,Vegas2} which is based on importance sampling. It samples
points from the probability distribution described by the function
$|f|$, so that the points are concentrated in the regions that make
the largest contribution to the integral. In the present version of
{\tt FIESTA} we use the FORTRAN code linked with the C interface to {\tt Mathematica}.  The VEGAS
routine requires an integrand as a reference to a function of an array
of the integration variables $x$, the number of repetitions and the number
of sampling points.

The {\tt Mathematica} part provides the integrand in a form of a long
symbolic algebraic expression consisting of the integration
variables, powers and logarithm\footnote{For the {\tt
  IfCut} method (see Appendix~\ref{D}) the {\tt if-then-else}
construction is also used, in the form {\tt if(condition)(first
branch)(second branch)}.}, e.g.:
\begin{verbatim}
(1+x[1])*p[x[1]+x[1]*x[2],-3]+ l[x[3]]/33
\end{verbatim}
where \verb|p[x[1]+x[1]*x[2],-3]| stands for $(x[1]+x[1]*x[2])^{-3}$ and
\verb|l[x[3]]| for $\log(x[3])$.

The common approach is based on generating the corresponding C (or
FORTRAN) code, compiling it, linking with some integration routine
(VEGAS in our case), running the resulting program and reading back
the answer. However, in our case this approach does not work properly:
it is specific to the sector decomposition that there can
be many thousands of sectors. This leads to a big overhead for
loading a compiler, a linker, writing the results to a file, loading
and starting the resulting program and reading the results back.

Next, the size of the corresponding expressions to be integrated may
vary from several bytes to hundreds of megabytes. Unfortunately, all
of the existing C or FORTRAN compilers are strongly nonlinear with respect
to the length of the input expression.
The reason is that these compilers usually use 
the so-called AST (Abstract Syntax Tree, see
\cite{dargonDook}) which is a tree representation of the syntax:
 each node represents constants
or variables (leaves) and operators or statements (inner nodes). The
complexity of this tree grows rapidly with the size of the incoming
expression.  In practice, it is nearly
impossible to compile an expression of more then a few megabytes
length.

In order to compile a huge expression one can use some
stack-based programming language like FORTH \cite{FORTH} 
but the performance of programs written on these languages is usually 
not so good as the performance of C or FORTRAN programs.

One of the standard approaches consists in optimizing the algebraic expression
by means of some CAS tool like the
Maple package ``CodeGeneration'' (formerly ``codegen'')\cite{Maple}
and compiling the resulting C (or FORTRAN) code without
optimization. This cannot solve our problem since the number of
expressions is too big, anyway, and the resulting (optimized) code
sometimes is still rather large. Moreover, the time the CAS spends to
generate such an optimized code is too large. So we decided to
develop an interpreter which is able to evaluate a given expression
in a ``data-driven'' manner: a translator translates the
incoming expression into some internal representation\footnote{We
use the triple form, see \cite{dargonDook}.} and then an
interpreter evaluates the expression many times for various input
data. It is worth noting that no intermediate
AST is
generated, only the linear sequence of triples.
This permits our
translator to be almost linear w.r.t. the length of the incoming expression.

\subsection{Translation}
The incoming expression consists of (many) terms. Each term is a
product of a coefficient and a number of integration variables (possibly zero) in some powers,
some prescribed functions of these variables and subexpressions of the same structure
(recursively). The translator converts this expression into a sequence
of triples, being ``an operation'', ``first operand'' and ``second operand''; e.g. the
expression \verb|1+x[1]+x[1]*x[3]|
will be represented as a following sequence of triples:
\begin{verbatim}
1: '+', 1, x[1]
2: '*', x[1], x[3]
3: '+', ^1, ^2
\end{verbatim}
where \^{}1 is a reference to the result of the first triple evaluation, \^{}2
to the second one. A term may involve some functions and
subexpressions, e.g. \\
\verb|x[1]+p[x[1],-2]*(x[2]+x[3])| will be translated to
\begin{verbatim}
1: 'p', x[1], -2
2: '+', x[2], x[3]
3: '*', ^1, ^2
4: '+', x[1], ^3
\end{verbatim}

During translation, the subexpression optimization is performed in the
following way: the translator searches for each new triple in the
dictionary. If the triple has already occurred, the translator
does not put the triple into the resulting sequence,
referencing the old one instead. Note, the
translator input is already {\em sorted} since we get it from
{\tt Mathematica} so at least all coinciding monomials will be evaluated
only once.

This simple prefix optimization is surely not perfect. Let us
consider the following expression:
\begin{verbatim}
(1+x[1]+x[2]*x[4])*(1+x[2]*x[4]+x[2]*x[3]*x[4])
\end{verbatim}

The terms are sorted but the subexpressions \verb|1+x[2]*x[4]| in
brackets will not be recognized, and \verb|x[2]*x[4]| in
\verb|x[2]*x[3]*x[4]| also will not be identified. Of course, we might have
implemented all possible combinations, but the resulting time would
grow factorially w.r.t. the length of the input string, which is unacceptable
for us. Hence we implemented the following strategy:

Let us consider a term. Let us assume it has an implicit coefficient 1 in
front.  We consider every factor as a ``commuting diad'' of a form
``operation'', ``operand'', e.g. \verb|3*x[1]*x[3]/x[2]| will be
considered as a set of the following commuting diads:
\begin{verbatim}
'*', 3
'*', x[1]
'*', x[3]
'/', x[2]
\end{verbatim}
Now we can re-order all these diads without affecting the result.
We try to evaluate such a sequence in some ``native'' order,
e.g. in the above example this would be like the following:
\begin{verbatim}
'*', x[1]
'/', x[2]
'*', x[3]
'*', 3
\end{verbatim}
We put the coefficient to the end, so terms differing by a coefficient
will be evaluated only once.
It is also worth noting that in each subexpression {\tt Mathematica} already
sums up all similar terms.

The same idea works out (even better) for summands, the only
difference is the implicit initial value 0 instead of 1.

This works well, reducing the number of generated triples by an order
of magnitude. The translation procedure is in the worst case quadratic
w.r.t. to the length of the input string, in average it is a bit worse
than linear. Typically, in a 3GHz Xeon processor the translator spends
about 30 seconds to translate a gigabyte expression,
which is incomparably less than the integration time.

\subsection{Evaluation}
\label{EE}
The sequence of the translated triples is completely platform
independent, and it is rather straightforward to generate (directly to
the memory) the binary executable code for any existing architecture,
and execute it. For the moment, we've  restricted ourselves by developing
the interpreter suitable for every platform.

First of all, the triple sequence we have from the translation stage, is translated into the array of quadrples:
``operation'', ``operand 1'', ``operand 2'' and ``result''. The fields ``operand
1'' and ``operand 2'' are pointers to some memory addresses.  The ``result'' field is
either the result of the quadruple evaluation (\verb|double| type in C), or the
address of the memory cell in which the result must be stored. The
point is that the size of the address field on 64 bit platforms is the
same as the size of \verb|double|, and the usage of indirect references is
rather slow, so it seems to be reasonable to use a new memory cell for
each intermediate result.

However, on a modern CPU such an approach appears to be completely wrong.
Modern processors have a big write-back cache memory (megabytes), and
provided the same memory address is permanently updated it is never
written to RAM but stays in the cache. On the other hand, if we write intermediate
results every time to a new address, all this data is sooner or later
written to RAM (after the cache exhausted). That is why
for ``small'' jobs, when all intermediate results fit into the
processor cache, the interpreter stores results directly to the
``result'' field of a quadruple, while for ``large-scale'' jobs the
translator creates some buffers and provides the quadruples with re-usable
addresses of elements of these buffers.

The algorithm switches from the ``small'' to ``large'' model when the
number of generated triples reaches some threshold which strongly
depends on the size of the processor L2 (L3) cache per active core and
it is a subject for tuning for every specific architecture. The
corresponding value is hard-coded, it can be changed in the file
scanner.h, the macro \\ INDIRECT\_ADDRESSING\_THRESHOLD.

The complete structure of the C-part is as follows.  The routine
\verb|AddString()| builds the integrand step by step collecting the
incoming lines. Every moment the process may be canceled by invoking
the function \verb|ClearString()|.  After (almost) all of the lines are
collected, the routine \verb|Integrate()| adds the rest and invokes
the translator. After the incoming expression is translated into the
quadruple array the routine \verb|Integrate()| invokes
the VEGAS routine for 5 repetitions with 10000 sampling points in
order to ``warmup'' the grid. Finally, the VEGAS routine is invoked by
the \verb|Integrate()| routine for 15 repetitions with 100000 sampling
points for the real integration\footnote{This is the default behavior; the exact number of sampling points
can be set by the {\tt VegasSettings} option, see Section~\ref{syntax}.}.

\section{Integration of sums vs sums of integrations}
\label{F}
The current version of {\tt FIESTA} integrates in each sector separately
(each sector, not just each primary sector). The results are summed up and the
error estimate is calculated with a mean-square norm. This approach might
encounter a natural suspicion: the errors add up and might result in an absurd result.
However practice shows that the real error of this method is normally only about 20 percent
greater that the error of the other approach --- integrating the sum of expressions.

The reason lies within the adaptable integration algorithms: if each of the terms is integrated
individually, the algorithm can adapt to all the peaks of the given term.
Hence the error estimate for the individual terms might be good enough.
However when one tries to evaluate the whole sum at once, the peaks and other peculiarities
of the different terms collide and the integration algorithm cannot adapt to all of them properly.

Another reason to integrate in each sector individually is obviously the economy of RAM.
And even more, this approach suits well for parallel environments due
to much better load balancing.

For the {\tt CIntegrate} interpreter short (but not {\em so-}short)
expressions are also preferable.

At first sight, long expressions could be optimized much better than the
short ones for the following reason: the longer the expression, the more common
subexpressions are encountered, and all the subexpressions should be evaluated only
once. But for a long expression the program must use the indirect
referencing for quadruple results (see Appendix~\ref{EE}) which can
slowdown the interpreter considerably.

In general, all these effects speedup the
evaluation up to an order of magnitude when each individual sector is
integrated independently on others.


\begin{thebibliography}{99}

\bibitem{theory}
N.N.~Bogoliubov and O.S.~Parasiuk, Izv. Akad. Nauk USSR 25 (1955) 429;
\\
N.N.~Bogoliubov and D.V.~Shirkov, Introduction to Theory of Quantized Fields, 3rd edition, Wiley, New York, 1983;
\\
K.~Hepp, Commun. Math. Phys., 2 (1966) 301;
\\
E.R.~Speer, J. Math. Phys.,  9 (1968) 1404;
\\
E.R.~Speer, Ann. Inst. H. Poincar\'e, 23 (1977) 1;
\\
P.~Breitenlohner and D.~Maison, Commun. Math. Phys., 52 (1977) 11, 39, 55;
\\
K.~Pohlmeyer, J. Math. Phys., 23, (1982) 2511;
\\
V.A.~Smirnov, Commun. Math. Phys., 134 (1990) 109;
\\
O.I.~Zavialov, Renormalized Quantum Field Theory, Kluwer Academic Publishers, Dodrecht, 1990;
\\
V.A.~Smirnov, Applied Asymptotic Expansions in Momenta and Masses, STMP 177, Springer, Berlin, Heidelberg, 2002.

\bibitem{BinothHeinric}
T.~Binoth and G.~Heinrich, Nucl. Phys. B, 585 (2000) 741 [hep-ph/0004013];
\\
T.~Binoth and G.~Heinrich, Nucl. Phys. B, 680 (2004) 375 [hep-ph/0305234];
\\
T.~Binoth and G.~Heinrich, Nucl. Phys. B, 693 (2004) 134 [hep-ph/0402265].

\bibitem{3box}
V. A. Smirnov, Phys. Lett. B567 (2003) 193 [hep-ph/0305142].

\bibitem{3loop}
T. Gehrmann, G. Heinrich, T. Huber, C. Studerus, Phys. Lett. B640 (2006) 252--259, [hep-ph/0607185].
\\
G. Heinrich, T. Huber and D. Maitre, Phys. Lett. B662 (2008) 344 [0711.3590].

\bibitem{4loop}
R. Boughezal and M. Czakon, Nucl. Phys. B755 (2006) 221, [hep-ph/0606232].
\\
T. Binoth and G. Heinrich, Nucl. Phys. B680 (2004) 375, [hep-ph/0305234].

\bibitem{Heinric}
G.~Heinrich, Int. J. of Modern Phys. A,  23 (2008) 10 [0803.4177];

\bibitem{BognerWeinzierl}
C.~Bogner and S.~Weinzierl, Comput. Phys. Commun., 178 (2008) 596 [0709.4092].
\\
C.~Bogner and S.~Weinzierl, [0806.4307].

\bibitem{Zeillinger}
D.Zeillinger, Enseign. Math. 52, 143 (2006).

\bibitem{Spivakovsky}
M.~Spivakovsky, Progr. Math. 36, 419 (1983).

\bibitem{dimreg}
G.~'t Hooft and M.~Veltman, Nucl.~Phys. B 44 (1972) 189;
\\
C.G.~Bollini and J.J.~Giambiagi, Nuovo Cim. 12 B (1972) 20.

\bibitem{Smirnov}
V.A.~Smirnov,
  ``Evaluating Feynman Integrals,''
  Springer Tracts Mod.\ Phys.\  211 (2004) 1;
\\
V.A.~Smirnov,
  ``Feynman integral calculus,''
Berlin, Germany: Springer (2006) 283 p.

\bibitem{SSS}
A.V.~Smirnov, V.A.~Smirnov and M.~Steinhauser, PoS(ACAT) (2007) 024 [0805.1871]
\\
A.V.~Smirnov, V.A.~Smirnov and M.~Steinhauser, to appear in Nucl.~Phys. B (Proc. Suppl.), [0807.0365]



\bibitem{Vegas1}
G. P. Lepage, J. Comp. Phys. 27, 192(1978).

\bibitem{Vegas2}
G. P. Lepage, the Cornell preprint CLNS-80/447,1980.

\bibitem{QLink}
QLink --- open-source program by A.V. Smirnov, http://qlink08.sourceforge.net

\bibitem{dargonDook}
 Alfred V. Aho {\em et al.}, ``Compilers: Principles, Techniques, and
 Tools (2nd Edition)'', Addison-Wesley, 2007.

\bibitem{FORTH}
http://forthlinks.com

\bibitem{Maple}
K. Geddes {\em et al.}, ``Maple 9 Advanced Programming Guide'', Maplesoft, 2003,\\
http://minerva.tau.ac.il/bsc/2/2130/docs/04/m9AdvancedProgrammingGuide.pdf,
p. 319

\end{thebibliography}
\end{document}